\documentclass[a4paper,12pt]{article}
\usepackage{theorem}
\usepackage{latexsym,amssymb,amsfonts,amsmath}
\usepackage[dvips]{graphicx}
\newtheorem{theorem}{Theorem}
\newtheorem{lemma}{Lemma}
\newtheorem{corollary}[theorem]{Corollary}

\newcommand{\bs}[1]{\boldsymbol{#1}}

\newcommand{\var}[2]{\varphi_{\boldsymbol{{#1}},{#2}}}

\title{{\Large {\bf Weak convergence of complex-valued measure for bi-product path space induced by quantum walk
}
}}

\author{ 
{\small 
Norio Konno,$^{1}$ 
\footnote{konno@ynu.ac.jp 
}\quad 
Etsuo Segawa$^{2}$ 
\footnote{e-segawa@m.tohoku.ac.jp
}\quad  
}\\ 
{\scriptsize $^{1}$ 
Department of Applied Mathematics, Faculty of Engineering, Yokohama National University
}\\
{\scriptsize Hodogaya, Yokohama 240-8501, Japan
} \\
{\scriptsize $^2$ 
Graduate School of Information Sciences, Tohoku University ,
}\\
{\scriptsize Aoba, Sendai, 980-8579, Japan
} \\
} 

\date{\empty }
\begin{document}
\maketitle

\par\noindent
\begin{small}
\par\noindent
{\bf Abstract}. 
In this paper, a complex-valued measure of bi-product path space induced by quantum walk is presented. 
In particular, we consider three types of conditional return paths in a power set of the bi-product path space 
(1) $\Lambda \times \Lambda $, (2) $\Lambda \times \Lambda'$ and (3) $\Lambda'\times \Lambda'$, 
where $\Lambda$ is the set of all $2n$-length ($n\in \mathbb{N}$) return paths and 
$\Lambda'(\subseteq \Lambda)$ is the set of all $2n$-length return paths going through $nx$ ($x\in [-1,1]$) at time $n$. 
We obtain asymptotic behaviors of the complex-valued measures for the situations (1)-(3) which 
imply two kinds of weak convergence theorems (Theorems 1 and 2). 
One of them suggests a weak limit of weak values. 

\end{small}

\setcounter{equation}{0}
\section{Introduction}
Let the set of all the $n$-truncated paths be $\Omega_n=\{-1,1\}^n$. 
Denote the coin space $\mathcal{H}_C$ spanned by choice of direction at each time step, that is, $\bs{e}_{-1}={}^T[1,0]$ and $\bs{e}_{1}={}^T[0,1]$. 
Let quantum coin on $\mathcal{H}_C$ be \[U=\begin{bmatrix} a & b \\ c & d \end{bmatrix}\in \mathrm{U}(2) \] 
with $abcd\neq 0$, where $\mathrm{U}(2)$ is the set of two-dimensional unitary matrices. 
Define weight of passage as $W: \Omega_n \to M_2(\mathbb{C})$ such that for $\xi=(\xi_n,\dots,\xi_1)\in \Omega_n$, 
\begin{equation} W(\xi)=P_{\xi_n}\cdots P_{\xi_1}\end{equation} 
with $P_j=\Pi_jU$, where $\Pi_j$ is projection onto $\bs{e}_j$. 
Here $M(2)$ is the set of all the complex-valued $2\times 2$ matrices. 
In this paper, we consider bi-product $n$-truncated path space $\Omega_n^2=\Omega_n\times \Omega_n$. 
The algebra of subsets of $\Omega_n^2$ is denoted by $\mathcal{F}_n=2^{\Omega_n^2}$. 
For fixed $\bs{\phi}\in \mathcal{H}_C$ with $||\phi||=1$ called initial coin state, 
we define $\varphi_{\bs{\phi},n}: \mathcal{F}_n \to \mathbb{C}$ by for any $A\in \mathcal{F}_n$, 
\begin{equation}\label{def} 
\var{\phi}{n}(A)
	=\left\langle \bs{\phi}, \sum_{(\xi,\eta)\in A} W(\xi)^\dagger \cdot W(\eta)\bs{\phi} \right\rangle. 
\end{equation}
If $A=\emptyset$, then $\varphi_{\bs{\phi},n}(A)\equiv 0$ for the convenience. 
We should remark that the map $\varphi_{n}$ expresses $\mathbb{C}$-valued measure on $\mathcal{F}_n$ in the following sense: 
for every $\bs{\phi}\in \mathcal{H}_C$ with $||\bs{\phi}||=1$, \\
\noindent \\
\noindent{\bf Property of $\varphi_{\phi,n}$}
\begin{enumerate}
\renewcommand{\theenumi}{\roman{enumi}}
\item\label{pe} For $A\in\mathcal{F}_n$, $\varphi_{\bs{\phi},n}(A)\in \mathbb{C}$. Furthermore, $\varphi_{n}(\Omega_n^2)=1$, 
\item\label{pi} For any $A_1,\dots,A_m\in \mathcal{F}_n$ with $A_i \cap A_j=\emptyset$ ($i\neq j$), 
	\[\var{\phi}{n}\left(\bigcup_{i=1}^m A_i\right)= \sum_{i=1}^m \var{\phi}{n}(A). \]
\end{enumerate}
In particular, for $\xi,\eta\in \Omega_n$, 
\[ \left(D\right)_{\xi,\eta}\equiv \var{\phi}{n}(\{(\xi,\eta)\})\]
is called the decoherence matrix starting from the initial coin state $\bs{\phi}$ which has been studied by \cite{G,GS,KS}. 
Moreover for any $A_0\in 2^{\Omega_n}$, 
$\nu_{n}(A_0)\equiv \varphi_{n}(A_0\times A_0)$ is called $q$-measure on $2^{\Omega_n}$ \cite{G,GS}. 

Let $\Omega^2=\Omega_n^2 \times \{-1,1\}^2 \times \{-1,1\}^2\times \cdots=(\{-1,1\}^2)^{\mathbb{N}}$. 
A subset $A\subset \Omega^2$ is a cylinder set if and only if there exist $n\in\{1,2,\dots\}$ and $B\in \mathcal{F}_n$ such that 
$A=B\times \{-1,1\}^2\times \{-1,1\}^2\times \cdots$. 
Denote $\mathcal{C}(\Omega^2)$ as the collection of all cylinder sets. 
From the unitarity of $U=P_1+P_{-1}$, we see that for $A\in\mathcal{F}_n$, 
\[ \var{\phi}{n+1}(A\times \{-1,1\}^2)
	=\left\langle \bs{\phi}, \sum_{(\xi,\eta)\in A} \left\{(P_1+P_{-1})W(\xi)\right\}^\dagger \cdot \left\{(P_1+P_{-1})W(\eta) \right\} \bs{\phi}\right\rangle=\var{\phi}{n}(A). \]
Thus if $A\in \mathcal{F}_n$, then 
\begin{equation}\label{po}
\var{\phi}{n+m}(A\times \{-1,1\}^{2m})=\var{\phi}{n}(A), 
\end{equation}
for any $m\geq 1$. 
Define $\varphi_{\bs{\phi}}: \mathcal{C}(\Omega^2)\to \mathbb{C}$ such that for any $A\in \mathcal{C}(\Omega^2)$ expressed 
by $A=B\times \{-1,1\}^2\times \{-1,1\}^2\times \cdots$ with $B\in \mathcal{F}_n$, 
\[ \varphi_{\bs{\phi}}(A)=\var{\phi}{n}(B). \]
Equation (\ref{po}) implies that if $B=B_1\times \{-1,1\}^{2(n-m)}$ with $B_1\in \mathcal{F}_m$ and $m\leq n$, then 
$\varphi_{\bs{\phi}}(A)=\var{\phi}{n}(B)=\varphi_{\bs{\phi},m}(B_1)$. So $\varphi_{\bs{\phi}}$ is well defined. 
Moreover, we easily find that $\varphi_{\bs{\phi}}$ satisfies both properties (\ref{pe}) and (\ref{pi}). 

For an initial coin state $\bs{\phi}$, the {\it usual} quantum walk at time $n$, $X_n^{(\bs{\phi})}$, originated by Gudder (1988)~\cite{G0}, 
is reexpressed by using $\varphi_{\bs{\phi}}$ as follows. 
For $j\in \mathbb{Z}$, and $n\in \mathbb{N}$, define $T_n^{(j)}\in \mathcal{C}(\Omega^2)$ as 
$T_n^{(j)}=\{(\xi,\eta)\in \Omega^2: \xi_1+\cdots+\xi_n=\eta_1+\cdots+\eta_n=j\}$, where $\Omega=\{-1,1\}^\mathbb{N}$. 
Indeed, we can check that $\varphi_{\bs{\phi}}(T_n^{(j)})\geq 0$, and $\sum_{j\in \mathbb{Z}}\varphi_{\bs{\phi}}(T_n^{(j)})=1$. 
The property (\ref{pe}) implies that 
\[ \varphi_{\bs{\phi}}\left(\Omega_n^2\setminus \bigcup_{j=-n}^n T_n^{(j)}\right)= 0. \]
Anyway, under the subalgebla $2^{\bigcup_{j=-n}^n T_n^{(j)}}\subset 2^{\Omega_n^2}$, 
the quantum walk at time $n$ is denoted by a {\it random} variable $X_n^{(\bs{\phi})}: \bigcup_{j=-n}^n T_n^{(j)}\to \{-n,-(n-1),\dots,n-1,n\}$. 
Here $X_n^{(\bs{\phi})}(\xi,\eta)=\xi_1+\cdots+\xi_n=\eta_1+\cdots+\eta_n$ has the following distribution: 
\[P(X_n^{(\bs{\phi})}=j)\equiv P\left(\left\{(\xi,\eta)\in \bigcup_{j=-n}^n T_n^{(j)}: X_n^{(\bs{\phi})}(\xi,\eta)=j\right\}\right)
	 =\varphi_{\bs{\phi}}(T_n^{(j)}).  \]
This is an equivalent expression for the definition of the usual quantum walk on $\mathbb{Z}$ which has been intensively studied by many researchers.  
Now using $\varphi_{\bs{\phi}}$, 
we can measure various kinds of cylinder sets including $T_n^{(j)}$ corresponding to the usual quantum walk. 
In the next section, we choose three kinds of $n$-truncated cylinder sets in $\mathcal{C}(\Omega^2)$ 
by using our measure $\varphi_{\bs{\phi}}$ and find their asymptotics for large $n$. 
\section{Results}
For $x,y\in \mathbb{R}$, 
define the set of all paths which go through the positions $nx$ and $ny$ at time $n$ and $2n$, respectively as follows:
\[ \Theta_{x|y}^{(n)}=\left\{ (\xi_1,\xi_2,\dots)\in \Omega: \frac{\xi_1+\cdots+\xi_n}{n}=x,\; \frac{\xi_{1}+\cdots+\xi_{2n}}{n}=y   \right\}. \]
Now we concentrate on $y=0$, and 
the following three cases with respect to 
the pair of $\Theta_x^{(n)}\times \Theta_y^{(n)}\in \mathcal{C}(\Omega^2)$, where $\Theta_x^{(n)}\equiv \Theta_{x|0}^{(n)}$ : 
\begin{enumerate}
\item $A_1^{(n)}\equiv \bigcup_{x\in\mathbb{R}}\bigcup_{y\in\mathbb{R}} \Theta_x^{(n)}\times \Theta_y^{(n)}$ case
\item $A_2^{(n)}(y)\equiv \bigcup_{x\in\mathbb{R}} \Theta_x^{(n)}\times \Theta_y^{(n)}$ with fixed $y\in \mathbb{R}$ case
\item $A_3^{(n)}(y)\equiv \Theta_y^{(n)}\times \Theta_y^{(n)}$ case
\end{enumerate}
Note that $A_1^{(n)}\supseteq A_2^{(n)}\supseteq A_3^{(n)}$. 
To explain the situations of $A_j^{(n)}$'s, we prepare two quantum walkers, walker 1 and walker 2, who produce the weight of path $W$. 
The measurement value is obtained by inner product of their weight of paths with an initial coin state (see Eq.~(\ref{def})).  
Both walkers in $A_1^{(n)}$ give weight of all the paths returning back to the origin at time $2n$. 
Walkers 1 and 2 in $A_3^{(n)}$ produce weight of every return path with length $2n$ restricted to passing the position $nx$ at time $n$. 
In $A_2^{(n)}$, despite of $A_1^{(n)}$ and $A_3^{(n)}$, 
the classes of return paths for two walkers are different: 
walker 1 is in the situation (1) while walker 2 is in the situation (3). 

The following theorem gives asymptotics of measurement value for each situation (1)-(3) by using $\var{\phi}{n}$. 
Define 
\begin{equation}\label{dkappa}
\mathcal{D}_\kappa=e^{i\kappa} \Pi_{-1}+\Pi_{1} \mathrm{\;\;with\;} \kappa=\mathrm{arg}(a)+\mathrm{arg}(c)-\mathrm{det}(U).
\end{equation} 
We use notation $a_n\sim b_n$ as $\lim_{n\to \infty}a_n/b_n=1$. 
\begin{lemma}\label{machiko}
Denote the Konno function $f_K(x;r)$ $(0<r<1)$ \cite{K,K1} by 
\[ f_K(x;r)=\frac{\sqrt{1-r^2}}{\pi (1-x^2)\sqrt{r^2-x^2}}\mathbf{1}_{\{|x|<r\}}(x), \]
where $\mathbf{1}_A(x)$ is the indicator function, that is, $\mathbf{1}_A(x)=1$, $(x\in A)$, $=0$, $(x\notin A)$.
Let the initial coin state be $\bs{\phi}_0$, 
and $\bs{\phi}_\kappa\equiv \mathcal{D}_\kappa \bs{\phi}_0$. Then we have for large $n$, 
\begin{enumerate}
\item \label{machiko1} Case (1)
\begin{equation}\label{pero1}
\varphi_{\bs{\phi}_0}(A_1^{(n)}) 
	\sim \frac{f_K(0;|a|)}{n}=\frac{|c|}{\pi|a| n}. 
\end{equation}
\item \label{machiko2} Case (2) \\ 
\begin{equation}
\sum_{j:j<ny}\varphi_{\bs{\phi}_0}\left( A_2^{(n)}(j/n) \right)
	\sim \mathbf{1}_{\{y>0\}}(y)\frac{f_K(0;|a|)}{n}=\mathbf{1}_{\{y>0\}}(y)\frac{|c|}{\pi |a|n}. 
\end{equation}
\item \label{machiko3} Case (3)
\begin{equation}\label{pero3}
\sum_{j\leq ny} \varphi_{\bs{\phi}_0}\left(A_3^{(n)}(j/n)\right)
	\sim \frac{|c|^2}{|a|^2 n}\int_{-\infty}^{y}\left(1+\langle \bs{\phi}_\kappa, C_0 \bs{\phi}_\kappa\rangle x\right)\frac{\mathbf{1}_{\{|x|<|a|\}}(x)}{\pi^2 (1-x^2)^2}dx,
\end{equation}
where \[C_0=\begin{bmatrix} 1 & -|c|/|a| \\ -|c|/|a| & -1 \end{bmatrix}.   \]
\end{enumerate}
\end{lemma}
Now we present a distribution function with respect to $q$-measure \cite{G,GS}. 
To do so, put 
\[ F_{n,\bs{\phi}_0}(x|y)
        \equiv \frac{\sum_{j\leq nx}\varphi_{\bs{\phi}_0}\left(\Theta_{(j/n)|y}^{(n)}\times \Theta_{(j/n)|y}^{(n)}\right)}
        		{\sum_{j\leq n} \varphi_{\bs{\phi}_0}\left(\Theta_{(j/n)|y}^{(n)}\times \Theta_{(j/n)|y}^{(n)}\right) }. \]
We can easily check that for fixed $y$, $F_{n,\bs{\phi}_0}(x|y)$ becomes a distribution function, that is, 
\begin{enumerate}
\renewcommand{\theenumi}{\alph{enumi}}
\item\label{one} $\lim_{x\to\infty}F_{n,\bs{\phi}_0}(x|y)=1$, $\lim_{x\to-\infty}F_{n,\bs{\phi}_0}(x|y)=0$, 
\item\label{two} for any $x\leq y$, $0\leq F_{n,\bs{\phi}_0}(x|z)\leq F_{n,\bs{\phi}_0}(y|z)\leq 1$. 
\end{enumerate}
The function $F_{n,\bs{\phi}_0}(x|y)$ corresponds to a {\it normalized} $q$-measure \cite{G,GS} restricted to the event $\bigcup_x \Theta_{x|y}^{(n)}$. 
Part (\ref{machiko3}) in Lemma \ref{machiko} leads the following theorem for $y=0$ case: 
\begin{theorem}
Assume that the initial coin state is $\bs{\phi}_0={}^T[\alpha,\beta]$. 
We consider the sequence $\{F_{n,\bs{\phi}_0}(x|0)\}_{n\geq 0}$. 
Let $Y_n$ be a random variable whose distribution function is $F_{n,\bs{\phi}_0}(x|0)$, that is, $P(Y_n\leq x)=F_{n,\bs{\phi}_0}(x|0)$. 
Then we have 
\begin{equation}
Y_n \Rightarrow Z,\;\;(n\to \infty)
\end{equation}
where $Z$ has the following density: 
\[ \nu_{\bs{\phi_0}}(x|0)=\frac{|c|^2}{|a|+|c|^2\log\sqrt{\frac{1+|a|}{1-|a|}}}
				\left[1-\left\{ (|\alpha|^2-|\beta|^2)+\frac{a\alpha\overline{b\beta}+\overline{a\alpha}b\beta}{|a|^2} \right\}x\right] \frac{\mathbf{1}_{\{|x|<|a|\}}(x)}{\pi^2 (1-x^2)^2} \]
Here ``$\Rightarrow$" means the weak convergence. 
\end{theorem}
Next, define 
$W^{(n)}_x=\bigcup_{y}\Theta_{y|x}^{(n)}$ and 
\[ \widehat{G}_{n,\bs{\phi}_0}(y|x)=\frac{\sum_{j\leq ny}\varphi_{\bs{\phi}_0}\left(W^{(n)}_x\times \Theta_{(j/n)|x}^{(n)}\right)}
	{\sum_{j\leq n} \varphi_{\bs{\phi}_0}\left(W^{(n)}_x\times \Theta_{(j/n)|x}^{(n)}\right)}. \]
The value $\widehat{G}_{n,\bs{\phi}_0}(y|x)$ satisfies the above condition (\ref{one}), but the condition (\ref{two}) is not ensured, that is, 
\[ \lim_{y\to\infty}\widehat{G}_{n,\bs{\phi}_0}(y|x)=1, \;\mathrm{and}\; \lim_{y\to-\infty}\widehat{G}_{n,\bs{\phi}_0}(y|x)=0, \]
while $\widehat{G}_{n,\bs{\phi}_0}(y|x)\in \mathbb{C}$ for $|y|<\infty$ in general. 
From Parts (\ref{machiko1}) and (\ref{machiko2}) in Theorem \ref{machiko}, we obtain an asymptotic behavior of 
the value $\widehat{G}_{n,\bs{\phi}_0}(x|0)$ which is deeply related to the weak value \cite{AAV,SH} as follows. 

Before we show the result, 
here we briefly give the definition of the weak value. 
We can see more detailed explanations and its interesting related works in \cite{Shikano} and its references. 
Let $\mathcal{H}$ be a Hilbert space and $U(t_2,t_1)$ be an evolution from time $t_1$ to $t_2$ on $\mathcal{H}$. 
For an observable $A$ and normalized states $\bs{\phi}_i,\bs{\phi}_f\in \mathcal{H}$, the weak value 
${}_{\phi_f}\langle A\rangle^w_{\phi_i}$ is defined by 
\begin{equation}\label{wv}
{}_{\phi_f}\langle A\rangle^w_{\phi_i}=\frac{\langle \bs{\phi}_f, U(t_f,t)AU(t,t_i)\bs{\phi}_i\rangle}{\langle \bs{\phi}_f,U(t_f,t_i)\bs{\phi}_i\rangle}.
\end{equation}
Here $\bs{\phi}_i$ and $\bs{\phi}_f$ are called pre-selected state and post-selected state, respectively. 

From now on, we take the Hilbert space $\mathcal{H}$ as $\bigoplus_{x\in \mathbb{Z}}\mathcal{H}_x$, where $\mathcal{H}_x$ is the two-dimensional Hilbert space 
spanned by left and right chiralities $\{\bs{e}_L,\bs{e}_R\}$. 
Let the canonical basis of $\mathcal{H}$ be denoted by $\{\bs{\delta}_x\otimes \bs{e}_L,\bs{\delta}_{x}\otimes \bs{e}_R; x\in \mathbb{Z}\}$. 
Put a permutation operator $S$ on $\mathcal{H}$ such that for $\bs{\delta}_x\otimes \bs{e}_J$ $(J\in \{L,R\})$, 
\[ S(\bs{\delta}_x\otimes \bs{e}_J)=\begin{cases} \bs{\delta}_{x+1}\otimes \bs{e}_R, & \text{$(J=R)$}, \\  \bs{\delta}_{x-1}\otimes \bs{e}_L, & \text{($J=L$).} \end{cases}\]
Define $E=SC$ be a unitary operator on $\mathcal{H}$, where $C=\sum_x \oplus U$. (Recall that $U$ is the two-dimensional unitary operator.) 
We consider the iteration of $E$ from the initial state $\bs{\Phi}_0=\bs{\delta}_0\otimes \bs{\phi}$ with $||\bs{\phi}||^2=1$:
\[\bs{\Phi}_0 \stackrel{E}{\mapsto}\bs{\Phi}_1\stackrel{E}{\mapsto}\bs{\Phi}_2\stackrel{E}{\mapsto}\cdots. \]
This is another equivalent expression for the quantum walk on $\mathbb{Z}$ with initial state $\bs{\Phi}_0$. 
Indeed, 
\[\big|\big| \Pi_j E^n \bs{\Phi}_0 \big|\big|^2=\varphi_{\bs{\phi}}(T_n^{(j)}), \]
where $\Pi_j$ is the projection onto $\mathcal{H}_j$. 

In particular, when we take for $t_1,t_2\in \mathbb{N}$, ${E}^{t_2-t_1}$ as $U(t_2,t_1)$ and $\Pi_{j}$ as the observable $A$, 
moreover $\bs{\Phi}_0$ as the pre-selected state and 
$\Pi_0\overline{U}^{t_f}\bs{\Phi}_0$ as the post-selected state in Eq.~(\ref{wv}) with $t_i=0$, $t=n$ and $t_f=2n$, then we have in this setting 
\begin{equation}
\sum_{j\leq ny} {}_{\phi_f}\langle A\rangle^w_{\phi_i}=\widehat{G}_{n,\bs{\phi}_0}(y|0). 
\end{equation}
This is a connection between our complex-valued measure and weak value. 
We find that the weak value weakly converges to the delta measure as follows. 
\begin{theorem}\label{weakvalue}
It is hold that for large $n$, 
\begin{align}
\lim_{n\to \infty}\widehat{G}_{n,\bs{\phi}_0}(y|0) =  \mathbf{1}_{\{y>0\}}(y). 
\end{align}
\end{theorem}
%
The physical meaning of Theorem \ref{weakvalue} remains as an interesting open problem. 
\section{Proof of Lemma \ref{machiko}}
Let $\Xi_n(j)=\sum_{\xi : \xi_n+\cdots+\xi_1=j }W(\xi)$ be weight of all the $n$-truncated passages arriving at $j$. 
Our proof is based on the stationary phase method: 
\begin{lemma}\label{spm}
Let $f(x)$ denote an $\mathbb{R}$-valued function on $[a,b]$ satisfying that there exists a unique $c\in [a,b]$ such that 
$f'(c)=0$ with $f''(c)\neq 0$. Then for large $n$, 
\begin{equation}\label{hellocafe}
\frac{1}{n}\sum_{j:\;an<j<bn}g(j/n)e^{inf(j/n)}\sim e^{i\mathrm{sgn}(f''(c))\pi/4}\sqrt{\frac{2\pi}{|f''(c)|n}}g(c)e^{inf(c)}+o\left(1/\sqrt{n}\right), 
\end{equation}
where for $y\in \mathbb{R}$, $\mathrm{sgn}(y)=1$, $(y>0)$, $=0$, $(y=0)$, $=-1$, $(y<0)$. 
\end{lemma}
At first we give the following key lemma whose proof is described in Appendix by using the stationary phase method: 
\begin{lemma}\label{acco}
Put $\mathbb{R}$-valued functions $k(x)$ and $\psi(x)$ $(x\in [-|a|,|a|])$ as 
\begin{align} 
e^{ik(x)} &= \frac{1}{|a|}\sqrt{\frac{|a|^2-x^2}{1-x^2}}+i\frac{|c|}{|a|}\frac{x}{\sqrt{1-x^2}},  \label{kx}\\
e^{i\psi(x)} &= \sqrt{\frac{|a|^2-x^2}{1-x^2}}+i\frac{|c|}{\sqrt{1-x^2}}. \label{psix}
\end{align}
For any $j\in \mathbb{Z}$ with $j=n x$ $(x\in [-1,1])$, we obtain 
\begin{multline}
\Xi_n(j)=\frac{1+(-1)^{n+j}}{2} e^{in\delta/2}\sqrt{\frac{2f_K(x;|a|)}{n}} \\
	\times \mathcal{D}_\kappa^\dagger \left(e^{i\pi/4}e^{in(\psi(x)-xk(x))}\Pi(x)
        +e^{-i\pi/4}e^{-in(\psi(x)-xk(x))}\overline{\Pi(x)}\right)\mathcal{D}_\kappa+o(1/\sqrt{n}), 
\end{multline}
where
\[\Pi(x)=\begin{bmatrix} |a|(1-x) & |c|x+i\sqrt{|a|^2-x^2} \\ |c|x-i\sqrt{|a|^2-x^2} & |a|(1+x) \end{bmatrix}.  \]
Here for $M\in M_2(\mathbb{C})$, $\bigg(\overline{M}\bigg)_{i,j}=\overline{(M)_{i,j}}$ for any $i,j\in \{1,2\}$. 
\end{lemma}

Before the proof of Lemma \ref{machiko}, we can confirm a consistency of the statement of the above lemma as follows. 
Recall that $X_n^{(\bs{\phi})}$ is a random variable determined by 
$P(X_n^{(\bs{\phi})}=j)=|| \Xi_n(j)\bs{\phi} ||^2$ with the initial coin state $\bs{\phi}=[\alpha,\beta]$ 
so called usual quantum walk. 
Then Lemma \ref{acco} and the Riemann-Lebesgue lemma imply the following corollary with respect to $X_n^{(\bs{\phi})}$: 
\begin{corollary} 
\[ \lim_{n\to \infty}P(X_n^{(\bs{\phi})}/n\leq x)
	=\int_{-\infty}^{x} \left\{1-\left(|\alpha|^2-|\beta|^2+\frac{2\mathrm{Re}[a\alpha \overline{b\beta}]}{|a|^2}\right)x \right\}f_K(x;|a|)dx. \]
\end{corollary}
This is consistent with results of \cite{K,K1}. 
Now we give the proof of Lemma~\ref{machiko} in the following:
\begin{enumerate}
\item {\it Proof of Case (1).} \\
Put $g(x)=\psi(x)-xk(x)$. 
We should remark that $L_n(x)\equiv \sum_{\xi\in \Theta^{(n)}_x}W(\xi)=\Xi_n(-nx)\Xi_n(nx)$. 
Note that $\sum_{j=-n}^n L_n(j/n)=\Xi_{2n}(0)$. Lemma \ref{acco} reduces to 
\begin{equation}\label{ikea}
e^{-in\delta/2}D_\kappa\Xi_{2n}(0)D_\kappa^\dagger 
	\sim \sqrt{\frac{f_K(0;|a|)}{n}} \left\{e^{i\frac{\pi}{4}}e^{2ing(0)}\Pi(0)+e^{-i\frac{\pi}{4}}e^{-2ing(0)}\overline{\Pi(0)}\right\}.  
\end{equation}
By using the fact that for every $x\in\mathbb{R}$, 
\begin{equation}\label{ikea0}
\Pi^2(x)=\Pi(x),\;\;\Pi(x)\overline{\Pi(-x)}=0,
\end{equation}
and Eq. (\ref{ikea}), we obtain 
\begin{align}
\varphi_{\bs{\phi}_0}(A_1^{(n)}) &= \sum_{i=-n}^{n} \sum_{j=-n}^{n} \langle L_n(i/n)\bs{\phi}_0, L_n(j/n)\bs{\phi}_0\rangle\\
	&= \langle \Xi_{2n}(0)\bs{\phi}_0, \Xi_{2n}(0)\bs{\phi}_0\rangle \\
        &\sim \frac{f_K(0;|a|)}{n}\left\langle\bs{\phi}_0,\left\{ \Pi(0)+\overline{\Pi(0)} \right\}\bs{\phi}_0\right\rangle = \frac{f_K(0;|a|)}{n}.
\end{align}
Then we complete the proof of case (1). 
It is consistent with the result of \cite{K2} which treats the Hadamard walk. 
\begin{flushright} $\square$ \end{flushright}
\item {\it Proof of Case (2).} \\
Using Eq.~(\ref{ikea0}), 
Lemma \ref{acco} implies that 
\begin{multline}\label{ikea2}
D_\kappa \Xi_n(-j)\Xi_n(j) D_\kappa^\dagger e^{-in\delta}
        \sim i\frac{1+(-1)^{n+j}}{2}\times  \frac{2f_K(x;|a|)}{n} \\
        	\times \left\{ e^{2ing(x)}\Pi(-x)\Pi(x)-e^{-2ing(x)}\overline{\Pi(-x)\Pi(x)} \right\},
\end{multline}
By Eq.~(\ref{ikea}), 
\begin{multline}\label{popopo}
e^{-in\delta}\sum_{j<ny}D_\kappa\Xi_n(-j)\Xi_n(j) D_\kappa^\dagger
	\\ \sim \frac{2i}{n}\sum_{j<ny}f_K(j/n;|a|)\bigg\{ e^{2ing(j/n)}\Pi(-j/n)\Pi(j/n) 
        	-e^{-2ing(j/n)}\overline{\Pi(-j/n)\Pi(j/n)} \bigg\}
\end{multline}
Now we consider the solution for $g'(x)=\psi'(x)-k(x)-xk'(x)=0$. 
Equations~(\ref{theta})-(\ref{renon}) in Appendix imply that $\psi(x)=\theta(k(x))$ and $k(x)$ is the unique solution for 
\begin{equation}\label{hk}  h(k)=\partial\theta(k)/\partial k=x \end{equation}
on $k\in [-\pi/2,\pi/2]$, 
where $\cos\theta(k)=|a|\cos k$ with $\sin \theta(k)\geq 0$. 
So we have
\[ \frac{\partial \psi(x)}{\partial x}=\frac{\partial \theta(k(x))}{\partial x}=xk'(x). \]
Then  we obtain 
\begin{equation}\label{g'k}
g'(x)=-k(x).
\end{equation}
On the other hand, differentiating both sides of Eq.~(\ref{hk}) with respect to $x$ implies 
\[ \frac{\partial}{\partial x}\left( \frac{\partial \theta(k)}{\partial k} \right)
	=\frac{\partial k}{\partial x}\left( \frac{\partial^2 \theta(k)}{\partial k^2} \right)=1. \]
Then Eq.~(\ref{g'k}) gives 
\begin{equation}\label{k'f_K} k'(x)=\frac{1}{\partial^2 \theta(k)/\partial k^2}\bigg|_{k=k(x)}=\pi f_K(x;|a|).  \end{equation}
Thus, $g'(x)=0$ if and only if $k(x)=0$, which implies $e^{ik(x)}=1$. Therefore by definition of $k(x)$ (see Eq.~(\ref{kx})), 
$x=0$ is the unique solution for $g'(x)=0$. 
Moreover Eqs.~(\ref{g'k}) and (\ref{k'f_K}) give $g''(x)=-k'(x)=-\pi f_K(x;|a|)$, which implies $g''(0)=-\pi f_K(0;|a|)$. 
So applying the stationary phase method described in Lemma \ref{spm} to Eq.~(\ref{popopo}), we obtain 
\begin{multline}\label{garalley} 
e^{-in\delta}\sum_{j<ny}D_\kappa\Xi_n(-j)\Xi_n(j) D_\kappa^\dagger \\
	\sim i\mathbf{1}_{\{y>0\}}(y)\left(e^{2in\psi(0)}e^{-i\pi/4}\Pi(0)-e^{-2in\psi(0)}e^{i\pi/4}\overline{\Pi(0)}\right)\sqrt{\frac{f_K(0;|a|)}{n}}. 
\end{multline}

Combining Eq.~(\ref{garalley}) with Eq.~(\ref{ikea}), we arrive at 
\begin{align}
\varphi_{\bs{\phi}_0}(A_2^{(n)}(y)) &= \sum_{j:j<ny} \left\langle \Xi_{2n}(0)\bs{\phi}_0, \Xi_n(-j)\Xi_n(j)\bs{\phi}_0\right\rangle 
	\sim \mathbf{1}_{\{y>0\}}(y)\frac{f_K(0;|a|)}{n}. \label{ikea3}
\end{align}
So we complete the proof. 
\begin{flushright} $\square$ \end{flushright}
\item {\it Proof of Case (3).} \\
Remark that 
\begin{equation}
\sum_{j\leq ny}\varphi_{\bs{\phi}_0}(A_3^{(n)}(j/n)) 
	= \sum_{j\leq ny}\langle L_n(j/n)\bs{\phi}_0, L_n(j/n)\bs{\phi}_0 \rangle.
\end{equation}
On the other hand, using the relations of $\Pi(x)$ described by Eq.~(\ref{ikea0}), 
Eq.~(\ref{ikea2}) gives 
\begin{equation}
L^\dagger_n (x)\cdot L_n (x)\sim \frac{|c|^2}{n^2|a|^2}(I+C_0x)\frac{\mathbf{1}_{\{|x|<|a|\}}(x)}{\pi^2 (1-x^2)^2}
\end{equation}
which leads the desired conclusion of case (3). 
\begin{flushright} $\square$ \end{flushright}
\end{enumerate}

\par
\
\par\noindent
\noindent
{\bf Acknowledgments.}
NK acknowledges financial support of the Grant-in-Aid for Scientific Research (C) of Japan Society for
the Promotion of Science (Grant No. 21540116). 
\par
\
\par

\begin{small}
\bibliographystyle{jplain}

\end{small}
\noindent \\
\begin{appendix}
\section{Proof of Lemma \ref{acco}}
We take the spatial Fourier transform of the weight of path $\Xi_n(j)$ such that 
\[ \widehat{\Xi}_n(k)=\sum_{j\in \mathbb{Z}}\Xi_n(j)e^{ijk}. \]
From the recurrence relation $\Xi_{n+1}(j)=Q\Xi_{n}(j-1)+P\Xi_{n}(j+1)$, we obtain
\[  \widehat{\Xi}_n(k)=\left(e^{ik}Q+e^{-ik}P\right)^n. \]
The eigenvalues and their corresponding normalized eigenvectors are expressed by 
$\lambda_{m}(k+\tau)$ and $\bs{v}_{m}(k+\tau)$, $(m\in\{0,1\})$, where 
\begin{align}\label{theta}
\lambda_{m}(k) &= e^{i\delta/2}\cdot e^{i(-1)^m \theta(k)}, \\
\bs{v}_{m}(k) &= \frac{1}{\sqrt{2\{1-|a|\cos[(-1)^m\theta(k)-k]\}}}D_\kappa^\dagger  
	\begin{bmatrix} |c| \\ |a|-e^{i((-1)^m \theta(k)-k)} \end{bmatrix},
\end{align}
where $\tau=\delta/2-\mathrm{arg}(a)$ and $D_\kappa$ is defined in Eq.~(\ref{dkappa}). 
Here $\cos \theta(k)=|a|\cos k$ with $\sin \theta(k)\geq 0$ and $\delta=\mathrm{arg}(\mathrm{det}(U))$. 
By the Fourier inversion theorem, we obtain for any $\gamma\in \mathbb{R}$, 
\begin{align}
\Xi_n(j) &= \int_\gamma^{2\pi+\gamma} \widehat{\Xi}_n(k)e^{-ijk} \frac{dk}{2\pi}, \notag \\
	&= e^{in\delta/2} \sum_{m\in \{0,1\}} 
        	\int_{\gamma+\tau}^{2\pi+\gamma+\tau} e^{in((-1)^m\theta(k)-xk)} \bs{v}_m (k)\bs{v}_m (k)^\dagger \frac{dk}{2\pi}, \label{weight}
\end{align}
where $x=j/n$. We choose an arbitrary parameter $\gamma$ as $-\tau-\pi/2$. 
From now on we apply the stationary phase method in Lemma~\ref{spm} to Eq.~(\ref{weight}). 
Put $f_m(k)= (-1)^m \theta(k)-xk$, $(m\in \{0,1\})$ as $\mathbb{R}$-valued function on $[-\pi/2,3\pi/2)$. 
The solution for 
$\partial f_m(k)/\partial k=0$ is given by 
\begin{equation}\label{renon}(-1)^m h(k)=x, \end{equation}
where $h(k)=\partial \theta(k)/\partial k$. In the following, we consider $m=0$ case. 
The definition of $\theta(k)$ gives 
\[h(k)=\frac{|a|\sin k}{\sqrt{1-|a|^2\cos^2 k}}. \] 
The solutions for $h'(k)=0$ in $[-\pi/2,3\pi/2)$ are $\pm \pi/2$. 
We denote $h_{\pm}(k)$ with $h(k)=h_+(k)+h_-(k)$ so that $h_+'(k)>0$ and $h_-(k)\leq 0$, 
as the function on $K_+=[-\pi/2,\pi/2)$ and $K_-=[\pi/2,3\pi/2)$, respectively. 
To apply the stationary phase method, we divide the integral in Eq.~(\ref{weight}) into the four parts as follows: 
\begin{equation}\label{takada}
e^{-in\delta/2}D_\kappa \Xi_n(j) D_\kappa^\dagger =
\sum_{m\in\{0,1\}}\sum_{\epsilon\in\{-,+\}}\int_{k\in K_\epsilon}e^{in\left((-1)^m\theta(k)-xk\right)} \bs{v}_m (k)\bs{v}_m (k)^\dagger \frac{dk}{2\pi}.
\end{equation}
An explicit expression for the solutions $k_{\pm}(x)$ for $h_\pm (k)=x$, respectively, are obtained as follows: 
\begin{align}\label{kimi1}
\cos k_{\pm}(x) &= \pm \frac{1}{|a|}\sqrt{\frac{|a|^2-x^2}{1-x^2}}, \\
\sin k_{\pm}(x) &= \frac{|c|}{|a|} \frac{x}{\sqrt{1-x^2}}.
\end{align}
Thus we have 
\begin{equation}\label{a0} 
\left|\frac{1}{\partial^2 f_0(k)/\partial k^2}\right|_{k=k_\pm(x)}
	= \left|\frac{1}{\partial h(k)/\partial k}\right|_{k=k_\pm(x)} =\pi f_K(x;|a|). 
\end{equation}
Moreover some algebraic computations give 
\begin{align}
\bs{v}_0 (k) \bs{v}_0 (k)^\dagger|_{k=k_{+}(x)} &= D_\kappa^\dagger \Pi(x) D_\kappa,\;\;
\bs{v}_0 (k) \bs{v}_0 (k)^\dagger|_{k=k_{-}(x)} = D_\kappa^\dagger \overline{\Pi(x)} D_\kappa, \notag \\
\bs{v}_1 (k) \bs{v}_1 (k)^\dagger|_{k=k_{+}(x)} &= D_\kappa^\dagger \overline{\Pi(-x)} D_\kappa,\;\;
\bs{v}_1 (k) \bs{v}_1 (k)^\dagger|_{k=k_{-}(x)} = D_\kappa^\dagger {\Pi(-x)} D_\kappa. \label{a1}
\end{align}
For the solutions of Eq.~(\ref{renon}) in $m=1$ case, 
we replace the parameter $x$ in the result on $m=0$ case given by the above discussion with $-x$. 
By putting $\psi(x)$ as $\psi(x)=\theta(k(x))$  with $k(x)\equiv k_+(x)$, 
note that $\theta(k_+(-x))=\psi(x)$, $\theta(k_-(x))=\pi-\psi(x)$, and $k_+(-x)=-k(x)$, $k_-(x)=-k(x)-\pi$. 
Inserting these relations and Eqs.~(\ref{a0}) and $(\ref{a1})$ into the formula in Lemma~{\ref{spm}} 
for each term $(\epsilon,m)\in\{(+,0),(+,1),(-,0),(-,1)\}$ in Eq.~(\ref{takada}), 
we have the desired conclusion. 
\begin{flushright} $\square$ \end{flushright}

\end{appendix}

\end{document}